# Localization of magnetic sources underground by a data adaptive tomographic scanner

**Paolo Mauriello**[1] and **Domenico Patella**[2]

[1]*Department of Science and Technology for Environment and Territory, University of Molise, Campobasso, Italy*
*(E-mail: mauriello@unimol.it)*

[2]*Department of Physical Sciences, University Federico II, Naples, Italy*
*(E-mail: patella@na.infn.it)*

## ABSTRACT

A tomography method is proposed to image magnetic anomaly sources buried below a non-flat ground surface, by developing the expression of the total power associated with a measured magnetic field. By discretising the integral relating a static magnetic field to its source terms, the total power can be written as a sum of crosscorrelation products between the magnetic field data set and the theoretical expression of the magnetic field generated by a source element of unitary strength. Then, applying Schwarz's inequality, an occurrence probability function is derived for imaging any distribution of magnetic anomaly sources in the subsurface. The tomographic procedure consists in scanning the half-space below the survey area by the unitary source and in computing the occurrence probability function at the nodes of a regular grid within the half-space. The grid values are finally contoured in order to single out the zones with high probability of occurrence of buried magnetic anomaly sources. Synthetic and field examples are discussed to test the resolution power of the proposed tomography.

## INTRODUCTION

Probability tomography is a new interpretation method, introduced in geophysical anomaly source imaging to face with the intrinsic uncertain nature of the field data inversion problem. Any measured field can, in fact, be thought of as due to a set of elementary sources, which can be grouped in different plausible configurations, all generating effects on the survey area compatible with the collected dataset, within the measurement accuracy and station density.

Probability tomography was originally developed for the self-potential method (Patella, 1997a,b) and then extended to geoelectrical (Mauriello, Monna and Patella, 1998; Mauriello and Patella, 1999a), natural-field electromagnetic induction (Mauriello and Patella, 1999b) and gravity (Mauriello and Patella, 2001a,b) methods. The aim of this paper is to extend the new approach also to the magnetic method for imaging the most probable location of the sources of anomalies of the steady geomagnetic field. A preliminary analysis of the physical properties of the static magnetic field is made in order to find a suitable gauge for the definition of a source occurrence probability function.

## STRUCTURE OF THE MAGNETIC FIELD

We assume a coordinate system with the $(x,y)$-plane at sea level and the $z$-axis positive upwards. Let $\mathbf{B}(\mathbf{r})$ be a static magnetic induction field, evaluated at a grid of points $\mathbf{r} \in S$, where $S$ is a portion of the earth's surface characterized by a topography function $z(x,y)$. Since $\mathbf{B}(\mathbf{r})$ is a divergence-free field, a vector potential $\mathbf{A}(\mathbf{r})$ exists such that

$$\mathbf{B}(\mathbf{r}) = \nabla \times \mathbf{A}(\mathbf{r}) . \tag{1}$$

With eq.1 into Maxwell equation $\nabla \times \mathbf{B}(\mathbf{r}) = \mu_0 \mathbf{J}(\mathbf{r})$, we obtain

$$\nabla[\nabla \cdot \mathbf{A}(\mathbf{r})] - \nabla^2 \mathbf{A}(\mathbf{r}) = \mu_0 \mathbf{J}(\mathbf{r}) , \tag{2}$$

where $\mathbf{J}(\mathbf{r})$ is the total current density including electric charge transport by conduction and convection and also non-dissipative Ampère currents due to magnetization. The freedom of field gauge transformation allows the condition $\nabla \cdot \mathbf{A}(\mathbf{r})=0$ to be postulated, in order to allow $\mathbf{A}(\mathbf{r})$ to satisfy Poisson's equation

$$\nabla^2 \mathbf{A}(\mathbf{r}) = -\mu_0 \mathbf{J}(\mathbf{r}) . \tag{3}$$

The general solution of eq.3 is (Jackson, 1975)

$$\mathbf{A}(\mathbf{r}) = \frac{\mu_0}{4\pi} \int_V \frac{\mathbf{J}(\mathbf{r}')}{|\mathbf{r} - \mathbf{r}'|} dV , \tag{4}$$

where $V$ is a volume containing inside all the magnetic sources and $\mathbf{r}' \in V$.

Using eq.1, the magnetic induction field can at last be written as

$$\mathbf{B}(\mathbf{r}) = \frac{\mu_0}{4\pi} \int_V \mathbf{J}(\mathbf{r}') \times \frac{\mathbf{r} - \mathbf{r}'}{\left| \mathbf{r} - \mathbf{r}' \right|^3} dV \ . \tag{5}$$

Eq.5 will be used later to introduce a **J**-*occurrence probability* (JOP) function, *i.e.* the probability which a current element obtains at $\mathbf{r}'$ as source of the $\mathbf{B}(\mathbf{r})$ field.

An equivalent approach for the description of $\mathbf{B}(\mathbf{r})$ can be developed using the total magnetization vector $\mathbf{M}(\mathbf{r})$. In fact, in a multipole expansion of $\mathbf{A}(\mathbf{r})$, the lowest order contribution due to a volume $dV$ about $\mathbf{r}'$ is the dipolar term (Jackson 1975)

$$d\mathbf{A}(\mathbf{r}) = \frac{\mu_0}{4\pi} \mathbf{M}(\mathbf{r}') \times \frac{(\mathbf{r} - \mathbf{r}')}{\left| \mathbf{r} - \mathbf{r}' \right|^3} dV \ , \tag{6}$$

where $\mathbf{M}(\mathbf{r}') = \frac{1}{2}\mathbf{r}' \times \mathbf{J}(\mathbf{r}')$ . Accordingly, $\mathbf{A}(\mathbf{r})$ can also be written as (Stratton 1941)

$$\mathbf{A}(\mathbf{r}) = \frac{\mu_0}{4\pi} \int_V \mathbf{M}(\mathbf{r}') \times \frac{(\mathbf{r} - \mathbf{r}')}{\left| \mathbf{r} - \mathbf{r}' \right|^3} dV \ , \tag{7}$$

and, using again eq.1, the magnetic induction field can at last be given also as

$$\mathbf{B}(\mathbf{r}) = \frac{\mu_0}{4\pi} \int_V \frac{3\mathbf{n}[\mathbf{n} \cdot \mathbf{M}(\mathbf{r}')] - \mathbf{M}(\mathbf{r}')}{\left| \mathbf{r} - \mathbf{r}' \right|^3} dV \ , \tag{8}$$

where $\mathbf{n}$ is the unit vector in the direction of $\mathbf{r} - \mathbf{r}'$.

Eq.8 will be used later to introduce a **M**-*occurrence probability* (MOP) function, *i.e.* the probability which an elementary magnetic dipole gets at $\mathbf{r}'$ as responsible of the $\mathbf{B}(\mathbf{r})$ field.

In geophysical exploration either the *z*-component or the modulus of the earth's magnetic field is usually measured and a scalar secondary field (the anomalous field) is evaluated in order to identify sources of local magnetic anomalies. More explicitely, in the case of the so called vertical field survey, a scalar anomalous field is obtained by subtracting from the measured vertical component of the earth's magnetic field the vertical component of the primary magnetic field (the earth's main magnetic field). Similarly, in the case of the so-called total field survey, the scalar anomalous field is obtained by subtracting from the measured modulus of the earth's magnetic field the modulus of the known primary magnetic field. Since the secondary field is always a very small fraction of the primary field, it can be readily shown that in the case of a total field survey the scalar secondary field is the projection of the vector secondary field onto the direction of the vector primary field (Blakely 1996). This direction can generally be assumed uniform within the areas normally considered in geophysical exploration (Parasnis 1997). Thus, from now on we will consider $\mathbf{B}(\mathbf{r})$ as the secondary vector field, and the scalar component of $\mathbf{B}(\mathbf{r})$ along any fixed direction as the object of study in the new tomographic approach.

## 3D J-OCCURRENCE PROBABILITY

Referring to eq.6, $\mathbf{B}(\mathbf{r})$ can be discretized as follows

$$\mathbf{B}(\mathbf{r}) = \sum_{q=1}^{Q} \mathbf{P}_q \times \frac{(\mathbf{r} - \mathbf{r}_q)}{\left| \mathbf{r} - \mathbf{r}_q \right|^3} \ , \tag{9}$$

where $Q$ is the total number of sources generating the secondary field. The *q*-th source is a small volume $\Delta V_q$ centered at $\mathbf{r}_q$, crossed by an electric current with local density $\mathbf{J}(\mathbf{r}_q)$. Its strength $\mathbf{P}_q$ is given as

$$\mathbf{P}_q = \frac{\mu_0}{4\pi} \mathbf{J}(\mathbf{r}_q) dV_q \ . \tag{10}$$

Indicating with $B_u(\mathbf{r})$ the component of $\mathbf{B}(\mathbf{r})$ along a generic direction, identified by the unit vector $\mathbf{u}$, we define the corresponding signal power $\Lambda_u$ over $S$ as

$$\Lambda_u = \int_{(S)} B_u^2(\mathbf{r}) dS \ , \tag{11}$$

which, taking into account eq.9, can be expanded as

$$\Lambda_u = \sum_{q=1}^{Q} \int_{(S)} B_u(\mathbf{r}) \left[ \mathbf{P}_q \times \frac{(\mathbf{r} - \mathbf{r}_q)}{\left| \mathbf{r} - \mathbf{r}_q \right|^3} \right] \cdot \mathbf{u} dS \ , \tag{12}$$

and then put in the form

$$\Lambda_u = \sum_{q=1}^{Q} \left[ \sum_{\nu=x,y,z} P_{q\nu} \int_{(S)} B_u(\mathbf{r}) \Im_{u\nu}(\mathbf{r} - \mathbf{r}_q) dS \right] . \tag{13}$$

The explicit expressions of the $\Im_{u\nu}(\mathbf{r} - \mathbf{r}_q)$ functions ($\nu=x,y,z$) appearing in eq.13 are

$$\Im_{ux}(\mathbf{r} - \mathbf{r}_q) = \frac{1}{\left| \mathbf{r} - \mathbf{r}_q \right|^3} \left[ (y - y_q)\mathbf{k} \cdot \mathbf{u} - (z - z_q)\mathbf{j} \cdot \mathbf{u} \right], \tag{14a}$$

$$\Im_{uy}(\mathbf{r} - \mathbf{r}_q) = \frac{1}{\left| \mathbf{r} - \mathbf{r}_q \right|^3} \left[ (z - z_q)\mathbf{i} \cdot \mathbf{u} - (x - x_q)\mathbf{k} \cdot \mathbf{u} \right], \tag{14b}$$

$$\Im_{uz}(\mathbf{r} - \mathbf{r}_q) = \frac{1}{\left| \mathbf{r} - \mathbf{r}_q \right|^3} \left[ (x - x_q)\mathbf{j} \cdot \mathbf{u} - (y - y_q)\mathbf{i} \cdot \mathbf{u} \right], \tag{14c}$$

where $\mathbf{i}, \mathbf{j}$ and $\mathbf{k}$ are the unit vectors of the reference *x*-, *y*- and *z*-axis, respectively.

We consider a generic integral in eq.13 and apply Schwarz's inequality, obtaining

$$\left[ \int_{(S)} B_u(\mathbf{r}) \mathfrak{I}_{uv}(\mathbf{r}-\mathbf{r}_q) dS \right]^2 \leq \int_{(S)} B_u^2(\mathbf{r}) dS \int_{(S)} \mathfrak{I}_{uv}^2(\mathbf{r}-\mathbf{r}_q) dS \ . \quad (15)$$

Assuming that the projection of $S$ onto the $(x,y)$-plane is a rectangle with sides $2X$ and $2Y$ along the $x$- and $y$-axis, respectively, the normalization rule for integrals extended over irregular domains (Smirnov, 1977) allows eq.15 to be written as

$$\left[ \int_{-X}^{X} \int_{-Y}^{Y} B_u(\mathbf{r}) \mathfrak{I}_{uv}(\mathbf{r}-\mathbf{r}_q) g(z) dx dy \right]^2$$
$$\leq \int_{-X}^{X} \int_{-Y}^{Y} B_u^2(\mathbf{r}) g(z) dx dy \int_{-X}^{X} \int_{-Y}^{Y} \mathfrak{I}_{uv}^2(\mathbf{r}-\mathbf{r}_q) g(z) dx dy \ , \quad (16)$$

where $g(z)$, which has the role of a *topographic surface regularization* factor, is

$$g(z) = \sqrt{1 + (\partial z/\partial x)^2 + (\partial z/\partial y)^2} \ . \quad (17)$$

Using the inequality 16, we can now define a three-component JOP function for 3D magnetic tomography in the most general case of non-flat topography, as

$$\eta_{uv}^{(J)}(\mathbf{r}_q) = C_{uv}^{(J)} \int_{-X}^{X} \int_{-Y}^{Y} B_u(\mathbf{r}) \mathfrak{I}_{uv}(\mathbf{r}-\mathbf{r}_q) g(z) dx dy \ , \quad (v=x,y,z),$$
$$(18)$$

with

$$C_{uv}^{(J)} = \left[ \int_{-X}^{X} \int_{-Y}^{Y} B_u^2(\mathbf{r}) g(z) dx dy \int_{-X}^{X} \int_{-Y}^{Y} \mathfrak{I}_{uv}^2(\mathbf{r}-\mathbf{r}_q) g(z) dx dy \right]^{-1/2} \ ,$$
$$(v=x,y,z). \quad (19)$$

Each $\eta_{uv}^{(J)}$ function satisfies the bounding condition $-1 \leq \eta_{uv}^{(J)}(\mathbf{r}_q) \leq +1$. At $\mathbf{r}_q$, three values of $\eta_{uv}^{(J)}$ ($v=x,y,z$) can be thus computed. Each value is interpreted as the probability, with which the homologous component $J_v$ ($v=x,y,z$) of a $\mathbf{J}$-field can at $\mathbf{r}_q$ be retained responsible of the measured component of the magnetic field. To clarify the meaning of probability attributed to $\eta_{uv}^{(J)}$, we recall that, in general, a probability measure $p$ is defined as a function assigning to every subset $E$ of a space of states $U$ a real number $p(E)$ which satisfies the conditions (Gnedenko, 1979)

$$p(E) \geq 0, \text{ for every } E, \quad (20)$$

if $E \cap F \equiv 0$, with $E,F \subset U$, $p(E \cup F) = p(E) + p(F)$, $\quad (21)$

$$p(U) = 1. \quad (22)$$

Considering that the presence of a source element at $\mathbf{r}_q$ is independent from the presence of another source element at another point, the function

$$\rho(\mathbf{r}_q) = \frac{|\eta(\mathbf{r}_q)|}{\int_{(V)} |\eta(\mathbf{r}_q)| dV} \quad (23)$$

can be defined as a probability density, as it allows a probability function to be deduced according to axioms 20, 21 and 22. Actually, $\eta(\mathbf{r}_q)$ differs from $\rho(\mathbf{r}_q)$ only for a constant multiplier and the explicit algebraic sign, which defines the direction of the corresponding vector component. This demonstration will tacitly apply also to all the other occurrence probability functions, which will be introduced later.

## 3D M-OCCURRENCE PROBABILITY

Referring to eq.8, $\mathbf{B}(\mathbf{r})$ can also be discretized as

$$\mathbf{B}(\mathbf{r}) = \sum_{q=1}^{Q} \frac{3\mathbf{n}_q(\mathbf{n}_q \cdot \mathbf{d}_q) - \mathbf{d}_q}{|\mathbf{r}-\mathbf{r}_q|^3} \ . \quad (24)$$

In eq.24, $\mathbf{n}_q$ is the unit vector in the direction $\mathbf{r}-\mathbf{r}_q$ and the generic element is a small volume $\Delta V_q$ centred at $\mathbf{r}_q$ with magnetization $\mathbf{M}(\mathbf{r}_q)$ and magnetic moment $\mathbf{d}_q$ given by

$$\mathbf{d}_q = \frac{\mu_0}{4\pi} \mathbf{M}(\mathbf{r}_q) dV_q \ . \quad (25)$$

The total power $\Lambda_u$, associated with $B_u(\mathbf{r})$, can now be expanded as

$$\Lambda_u = \sum_{q=1}^{Q} \left[ \sum_{v=x,y,z} d_{qv} \int_{(S)} B_v(\mathbf{r}) \mathfrak{R}_{uv}(\mathbf{r}-\mathbf{r}_q) dS \right] \ . \quad (26)$$

The explicit expressions of the $\mathfrak{R}_{uv}(\mathbf{r}-\mathbf{r}_q)$ functions ($v=x,y,z$) are given by

$$\mathfrak{R}_{ux}(\mathbf{r}-\mathbf{r}_q) = \frac{1}{|\mathbf{r}-\mathbf{r}_q|^3} [3n_{qx}(n_{qx}\mathbf{i}\cdot\mathbf{u} + n_{qy}\mathbf{j}\cdot\mathbf{u} + n_{qz}\mathbf{k}\cdot\mathbf{u}) - \mathbf{i}\cdot\mathbf{u}],$$
$$(27a)$$

$$\mathfrak{R}_{uy}(\mathbf{r}-\mathbf{r}_q) = \frac{1}{|\mathbf{r}-\mathbf{r}_q|^3} [3n_{qy}(n_{qx}\mathbf{i}\cdot\mathbf{u} + n_{qy}\mathbf{j}\cdot\mathbf{u} + n_{qz}\mathbf{k}\cdot\mathbf{u}) - \mathbf{j}\cdot\mathbf{u}],$$
$$(27b)$$

$$\mathfrak{R}_{uz}(\mathbf{r}-\mathbf{r}_q) = \frac{1}{|\mathbf{r}-\mathbf{r}_q|^3} [3n_{qz}(n_{qx}\mathbf{i}\cdot\mathbf{u} + n_{qy}\mathbf{j}\cdot\mathbf{u} + n_{qz}\mathbf{k}\cdot\mathbf{u}) - \mathbf{k}\cdot\mathbf{u}],$$
$$(27c)$$

where $n_{qv}$, with $v=x,y,z$, are the components of the unit vector $\mathbf{n}_q$.

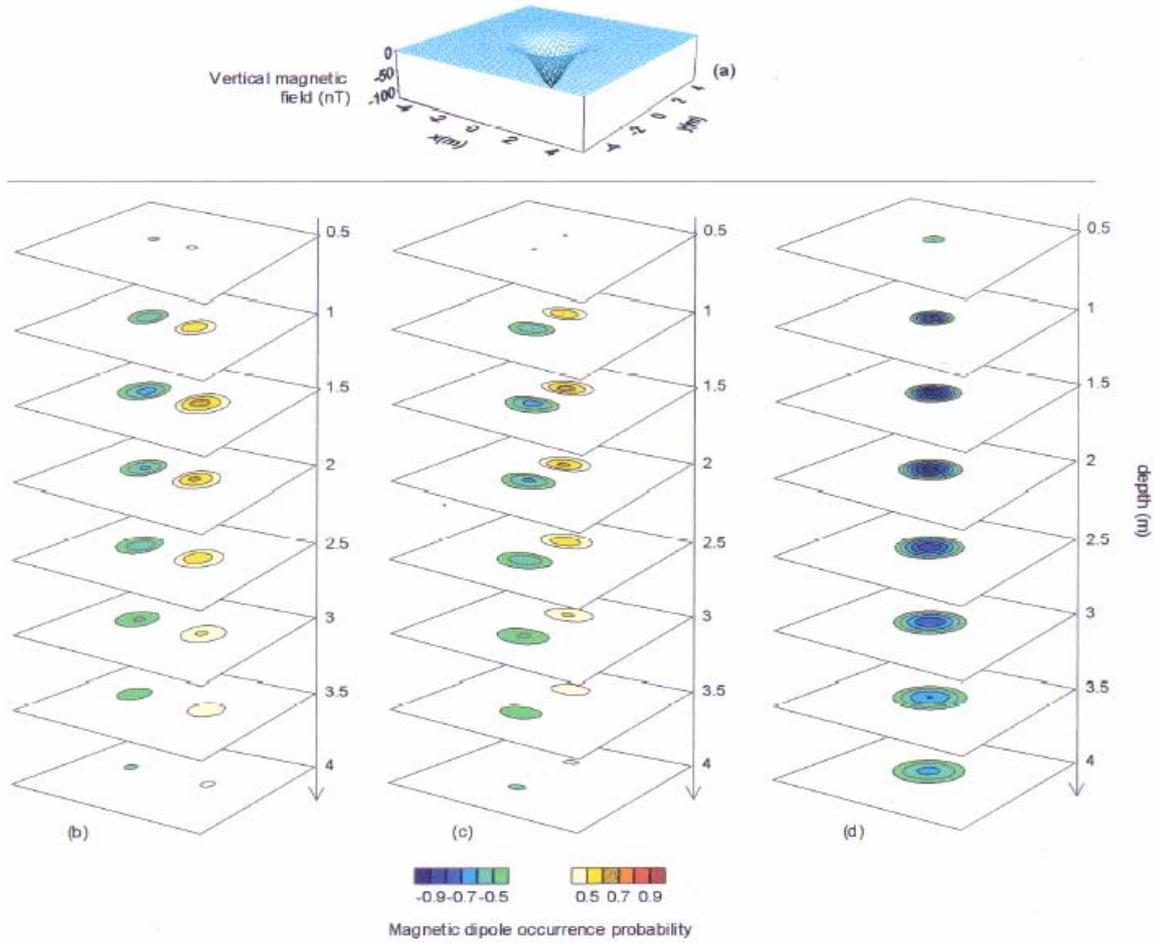

**Figure 1**    Probability tomography for a vertical magnetic dipole in the case of measurement of the *z*-component of the corresponding magnetic field. Synthetic surface map of $B_z$ (a), and 3D imaging of the magnetization occurrence probability function along the *x*-axis (b), *y*-axis (c) and *z*-axis (d).

Proceeding as before, we can now define a three-component MOP function as

$$\eta_{uv}^{(M)}(\mathbf{r}_q) = C_{uv}^{(M)} \int_{-X}^{X} \int_{-Y}^{Y} B_u(\mathbf{r}) \Re_{uv}(\mathbf{r} - \mathbf{r}_q) g(z) dx dy \ , \ (v{=}x,y,z),$$
$$(28)$$

where

$$C_{uv}^{(M)} = \left[ \int_{-X}^{X} \int_{-Y}^{Y} B_u^2(\mathbf{r}) g(z) dx dy \int_{-X}^{X} \int_{-Y}^{Y} \Re_{uv}^2(\mathbf{r} - \mathbf{r}_q) g(z) dx dy \right]^{-1/2},$$
$$(v{=}x,y,z). \qquad (29)$$

At every $\mathbf{r}_q$, three $\eta_{uv}^{(M)}$ values ($v{=}x,y,z$) can again be computed, each satisfying the bounding condition $-1 \le \eta_{uv}^{(M)}(\mathbf{r}_q) \le +1$. Each $\eta_{uv}^{(M)}$ value is interpreted as the probability by which the homologous component $M_v$ ($v{=}x,y,z$) of a **M**-field can be considered responsible for the measured component of the magnetic field.

## 3D PROBABILITY TOMOGRAPHY PROCEDURE

The 3D tomography procedure for imaging the sources of a magnetic field, measured on a generally uneven topography, consists in a reiterated computation code, involving the scanner functions $\Im_{uv}$ and $\Re_{uv}$ and the $B_u(\mathbf{r})$ field data set.

In practice, since we do not know the position of the real sources generating the anomalous magnetic field, we use a synthetic source of unitary strength to scan the *x,y,z* half-space below the surveyed area (tomospace), in order to search where the real sources can be located in a probabilistic sense. The scanning operation is made by computing the crosscorrelation integrals in eq.18 and eq.28 for each point $(x_q,y_q,z_q)$ of a regular grid within the tomospace. At each point, the value of each integral is interpreted as the occurrence probability of the relative magnetic source component, whose positive or negative orientation depends on whether it is $\eta{>}0$ or $\eta{<}0$. By scanning the tomospace using, *e.g.*, a sequence of horizontal slices spaced from each other by a constant depth interval, we can finally obtain a 3D image of the equivalent magnetic source distribution underground in a probabilistic sense.

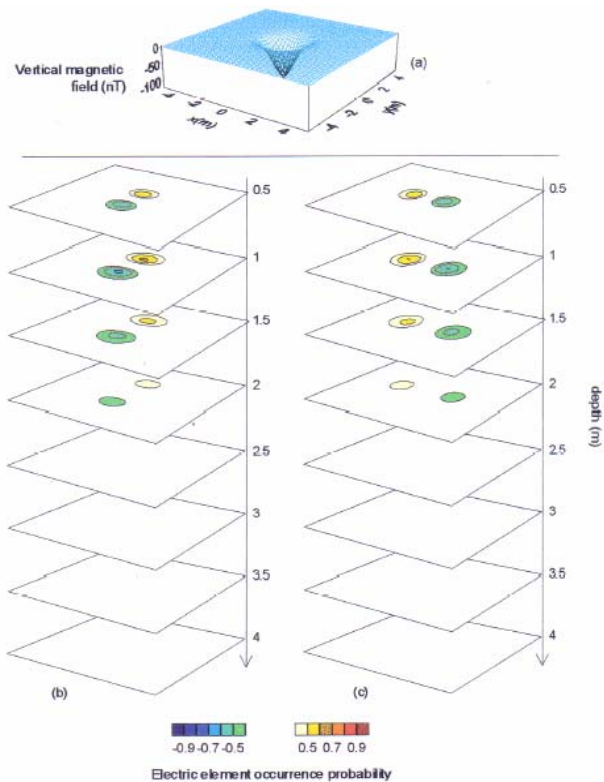

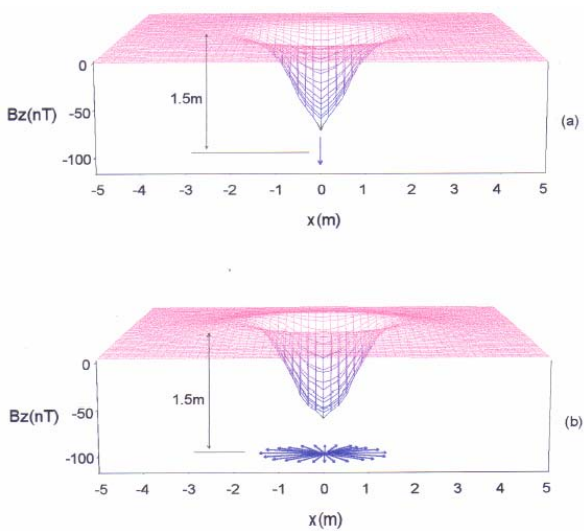

**Figure 2** Probability tomography for a vertical magnetic dipole in the case of measurement of the $z$-component of the magnetic field. Synthetic surface map of $B_z$ (a), and 3D imaging of the electric current occurrence probability function along the $x$-axis (b) and $y$-axis (c).

**Figure 3** Response of a distribution of dipoles equivalent to the vertical magnetic dipole in the case of measurement of the $z$-component of the geomagnetic field. The $B_z$ map due to the original vertical magnetic dipole (a), and to a radial sequence of horizontal dipoles (b).

In order to improve the filtering capability of the scanning procedure, for each $\mathbf{r}_q$ of the tomospace it is advisable to use varying sizes of the integration surface in eq.18 and eq.28. The smallest surface is the domain $[-X,X]\times[-Y,Y]$ wholly containing the surface trace of the magnetic response of the scanning element placed at $\mathbf{r}_q$. The greatest surface is, of course, the largest rectangle fitting to the whole survey area. The highest $\left|\eta(\mathbf{r}_q)\right|$ is then taken with its sign as the most appropriate source occurrence probability at $\mathbf{r}_q$.

## 3D SYNTHETIC EXAMPLES

In order to test the resolution power of the probability tomography, we present the results of the application to a synthetic case of magnetic dipole with three different dips. The dipole is centered at $(0,0,-1.5)$ *m* below a flat ground surface, with moment components $(0,0,-1)$ $Am^2$ (vertical dipole), $(1,0,0)$ $Am^2$ (horizontal dipole) and $(1/\sqrt{2},0,-1/\sqrt{2})$ $Am^2$ (45° downdipping dipole).

We have always considered, as initial dataset, the $z$-component of $\mathbf{B}(\mathbf{r})$, say $B_z(\mathbf{r})$ ($u\equiv z$). Using eqs.14a,b,c and eqs.27a,b,c, the three components of the $\Im_{zv}$ and $\Re_{zv}$ functions are given respectively by

$$\Im_{zx}(\mathbf{r}-\mathbf{r}_q)=\frac{y-y_q}{\left|\mathbf{r}-\mathbf{r}_q\right|^3}\,, \tag{30a}$$

$$\Im_{zy}(\mathbf{r}-\mathbf{r}_q)=\frac{x_q-x}{\left|\mathbf{r}-\mathbf{r}_q\right|^3}\,, \tag{30b}$$

$$\Im_{zz}(\mathbf{r}-\mathbf{r}_q)=0\,, \tag{30c}$$

and

$$\Re_{zx}(\mathbf{r}-\mathbf{r}_q)=\frac{3n_{qx}n_{qz}}{\left|\mathbf{r}-\mathbf{r}_q\right|^3}\,, \tag{31a}$$

$$\Re_{zy}(\mathbf{r}-\mathbf{r}_q)=\frac{3n_{qy}n_{qz}}{\left|\mathbf{r}-\mathbf{r}_q\right|^3}\,, \tag{31b}$$

$$\Re_{zz}(\mathbf{r}-\mathbf{r}_q)=\frac{3n_{qz}n_{qz}-1}{\left|\mathbf{r}-\mathbf{r}_q\right|^3}\,. \tag{31c}$$

Furthermore, in calculating the previous functions, we have always used the constant sampling step of 0.5 *m* along the three spacial directions. Finally, in plotting the tomographies, we have contoured only the values of the $\eta_{zv}^{(J)}$ and $\eta_{zv}^{(M)}$ functions ($v\equiv x,y,z$) exceeding in modulus 0.4, in order to not visualize secondary effects

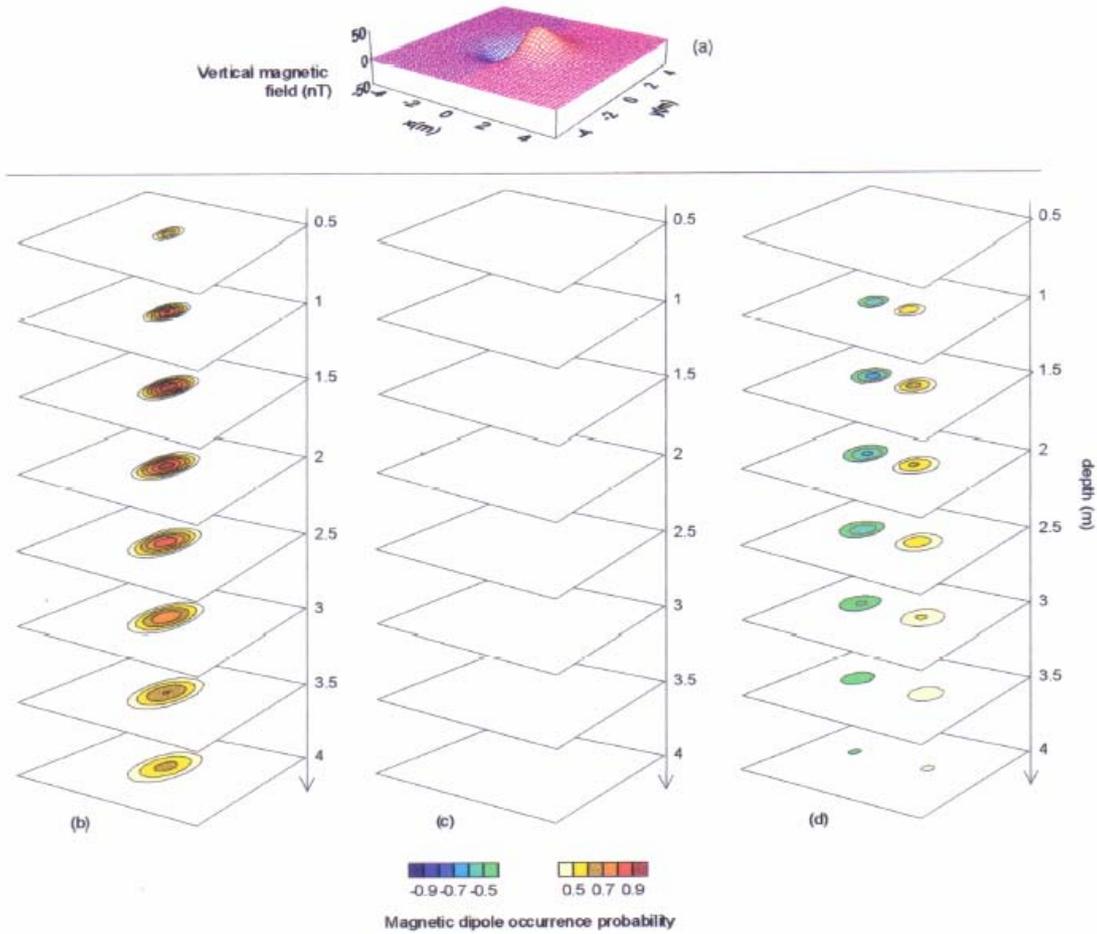

**Figure 4** Probability tomography for a horizontal magnetic dipole in the case of measurement of the *z*-component of the magnetic field. Synthetic surface map of $B_z$ (a), and 3D imaging of the magnetization occurrence probability function along the *x*-axis (b), *y*-axis (c) and *z*-axis (d).

due both to equivalent sources of no practical relevance and to numerical background noise.

### *The vertical magnetic dipole source*

The MOP and JOP tomographies are drawn in fig.1 and fig.2, respectively, where the topmost slice (a) is the simulated $B_z$ survey map due to the vertical dipole. In fig.1, the sequences of slices versus depth, reported in the columns (b), (c) and (d), show the behaviour of the MOP functions $\eta_{zx}^{(M)}$, $\eta_{zy}^{(M)}$ and $\eta_{zz}^{(M)}$, respectively. In fig.2, the two sequences in columns (b) and (c) show, instead, the behaviour of only the JOP functions $\eta_{zx}^{(J)}$ and $\eta_{zy}^{(J)}$; $\eta_{zz}^{(J)}$ cannot, in fact, be computed since $\mathfrak{I}_{zz}$ vanishes everywhere (see eq.30), as a vertical current element gives no contribution to the measured vertical component of the magnetic field.

Referring to the slice at -1.5 *m* of depth, we observe in fig.1d that $\eta_{zz}^{(M)}$ shows a minimum, which exactly outlines the position of the original downward pointing vertical dipole. Furthermore, figs.1b and 1c show two pairs of nuclei with opposite sign along the *x*- and *y*-axis, respectively. The highest absolute values of $\eta_{zx}^{(M)}$ and $\eta_{zy}^{(M)}$ appear at -1.5 *m* of depth, and are less than the modulus of the minimum of $\eta_{zz}^{(M)}$ ascribed to the original vertical dipole. These two pairs are interpreted as the signatures of an equivalent source consisting of a radial set of horizontal, outward oriented magnetization vectors. The $B_z$ map of fig.3a related to the original vertical dipole can, in fact, be accurately reproduced using, *e.g.*, a regular set of radial dipoles every 15°, as illustrated in fig.3b.

The $\eta_{zx}^{(J)}$ and $\eta_{zy}^{(J)}$ tomographies of figs.2b and 2c show two pairs of nuclei with opposite sign along the *y*- and *x*-axis, respectively. The highest absolute values of $\eta_{zx}^{(J)}$ and $\eta_{zy}^{(J)}$ appear at -1 *m* of depth. The two pairs are interpreted as traces of the topmost portions of the vertical current loops corresponding to the horizontal dipoles, appearing at -1.5 *m* in figs.1b and 1c. These traces completely mask the effect of the horizontal current loop equivalent to the original vertical dipole, which can ideally be imaged, joining together the less intense nuclei of $\eta_{zx}^{(J)}$ and $\eta_{zy}^{(J)}$ at -1.5 *m* of depth.

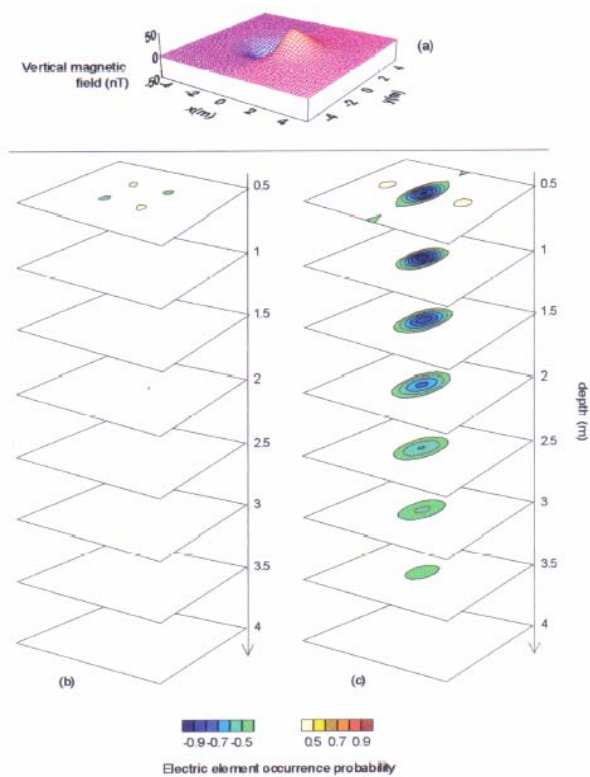

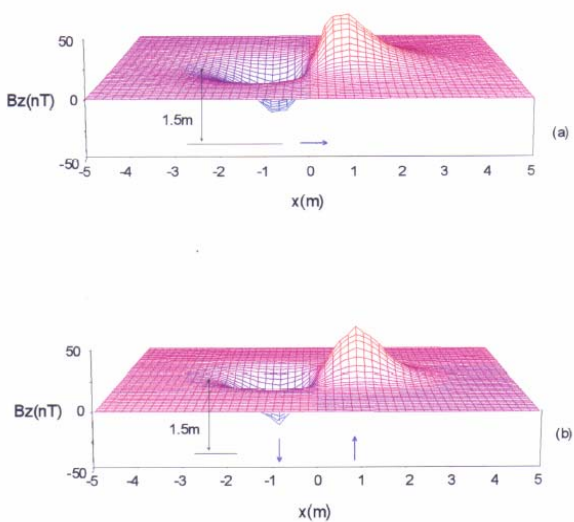

**Figure 5** Probability tomography for a horizontal magnetic dipole in the case of measurement of the *z*-component of the magnetic field. Synthetic surface map of $B_z$ (a), and 3D imaging of the electric current occurrence probability function along the *x*-axis (b) and *y*-axis (c).

**Figure 6** Response of dipole distributions equivalent to the horizontal magnetic dipole in the case of measurement of the *z*-component of the magnetic field. The $B_z$ map due to the original horizontal magnetic dipole (a), and to a pair of vertical dipoles with reverse polarity (b).

### The horizontal magnetic dipole source

The MOP and JOP tomographies are drawn in fig.4 and fig.5, respectively, where the topmost slice (a) is the simulated $B_z$ map due to the horizontal dipole. The sequences of slices in fig.4 and fig.5 have the same meaning as in fig.1 and fig.2, respectively.

In fig.4b, $\eta_{zx}^{(M)}$ gets the highest values at -1.5 *m* of depth within a nucleus, which correctly indicates the horizontal position of the original magnetization vector along the *x*-axis. The signature of the upper part of the equivalent vertical current loop is highlighted by the negative nucleus at -1 *m* of depth in fig.5c, where $\eta_{zy}^{(J)}$ shows the smallest values.

Fig.4d displays also a pair of nuclei with opposite sign. A pair of vertical dipoles with opposite polarity can be imaged at -1.5 *m* of depth, where $\eta_{zz}^{(M)}$ reaches the highest values. This additional pair, characterized by an occurrence probability lower than that observed in fig.4b, can again be interpreted as the signature of an equivalent source, whose conformity with the reference $B_z$ map, redrawn in fig.6a, is sketched in fig.6b. Of course, a pair of horizontal reverse current loops will correspond to such a pair of vertical dipoles. However, no traces of these loops appear along the *x*- and *y*-axis in figs.5b and 5c, since the relative JOP values are in modulus less than 0.4.

### The inclined magnetic dipole source

The MOP and JOP tomographies are given in fig.7 and fig.8, respectively, where the topmost slice (a) shows, as usual, the simulated $B_z$ field map due to the inclined dipole. The sequences of slices in fig.7 and fig.8 have the same meaning as in fig.1 and fig.2, respectively.

Fig.7 and fig.8 show clear transitional effects from the vertical (fig.1 and fig.2) to the horizontal dipole (fig.6 and fig.7). In fact, the effects due to the upper negative pole of the dipping dipole markedly dominate over the vanishing effects referred to the lower positive pole. In both the MOP and JOP plots, the distortion of the isolines appears as the consequence of a shift of the whole set of nuclei towards the negative part of the *x*-axis. Equivalently, the distortion can be thought of as the reaction of the magnetic field to an ideal rotation of the dipole from the vertical to the horizontal position, counterclockwise around a horizontal axis through its center, parallel to the *y*-axis.

The center of the original dipping dipole is exactly highlighted by the maximum of $\eta_{zx}^{(M)}$ in fig.7b and the minimum of $\eta_{zz}^{(M)}$ in fig.7d, both appearing at -1.5 *m* of depth. The original inclined magnetization vector can, in fact, be decomposed into a horizontal and a vertical magnetization vectors, contributing to the total effect as in fig.4b and fig.1b, respectively. There is, however, no direct indication about the dip angle, whose estimate can be done using standard rules on a profile obtained from the surface map through the dipole axis, knowing the depth of the dipole (Parasnis, 1997).

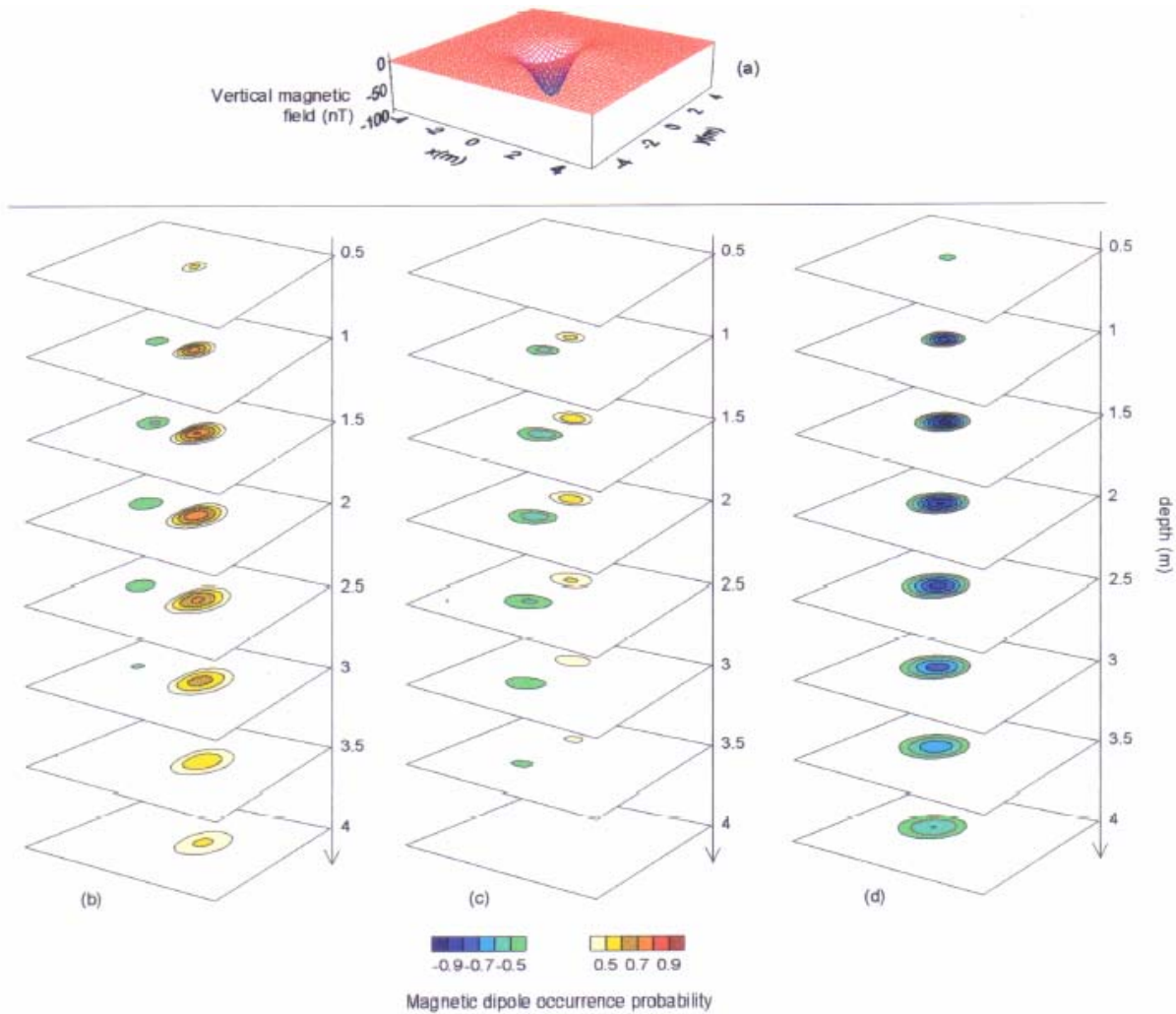

**Figure 7** Probability tomography for a 45° downward dipping magnetic dipole in the case of measurement of the *z*-component of the magnetic field. Synthetic surface map of $B_z$ (a), and 3D imaging of the magnetization occurrence probability function along the *x*-axis (b), *y*-axis (c) and *z*-axis (d).

### 3D FIELD EXAMPLES

To illustrate the applicability of the new 3D probability tomography imaging method to field cases, we discuss the results of two surveys performed in test sites where a reliable control of the solutions could be made. The first example relates to archaeology and the second one to volcanology.

#### *Application to archaeology*

A magnetic survey was done in the archaeological site of a Sabine Necropolis at Colle del Forno, 30 *km* north of Rome (Cammarano et al., 1998), in order to detect hypogeal *dromos-chamber* tombs, the typical geometry of which is shown in fig.9 (Santoro, 1977). A Geoscan FM36 fluxgate gradiometer was utilised to measure the gradient of the vertical magnetic component by a fixed vertical spacing of 50 *cm* between the two sensors. We consider here the results within an area of 10×10 $m^2$, where the presence of a tomb was already hypothesised by previous geoelectric surveys (Mauriello, Monna and Patella 1998; Mauriello and Patella 1999a).

Fig.10a and fig.11a show the experimental map of the difference between the vertical component of the earth's magnetic field at the height of 30 *cm* above the

8                              

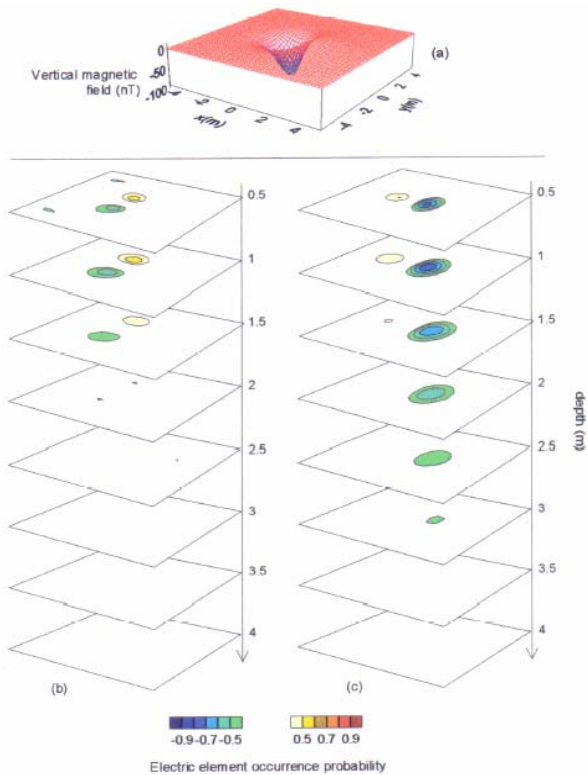

**Figure 8** Probability tomography for a 45° downward dipping magnetic dipole in the case of measurement of the *z*-component of the magnetic field. Synthetic surface map of $B_z$ (a), and 3D imaging of the electric current occurrence probability function along the *x*-axis (b) and *y*-axis (c).

ground and the vertical component at 80 *cm* above the ground which is assumed to approximate the primary earth's magnetic field. A composite magnetic minimum appears in the map exactly where a pair of resistive and conductive anomalies was previously detected by the geoelectrical survey, ascribed, respectively, to a well preserved chamber and to a corridor (*dromos*) entirely filled with sediments (Mauriello and Patella 1999a).

The sequences of slices in figs.10b, 10c, 10d show the behaviour of the MOP functions $\eta_{zx}^{(M)}$, $\eta_{zy}^{(M)}$, $\eta_{zz}^{(M)}$, respectively, whereas those in figs.11b, 11c show the behaviour of the JOP functions $\eta_{zx}^{(J)}$, $\eta_{zy}^{(J)}$, respectively. We observe in both the MOP and the JOP tomographies three separated features: the first one is very shallow at around -0.5 *m* of depth, the second one ranges between -1.5 *m* and -2.0 *m* of depth, and the last one lies at not less than -4 *m* of depth. Leaving out the first feature, which can easily be ascribed to the corridor before the tomb, the sequences of nuclei belonging to the other two features show in both tomographies of fig.10 and fig.11 a close similarity with those of fig.7 and fig.8, respectively, due to an inclined dipole. These sources would act in such a way as to locally lower the primary earth's magnetic field. Inclined dipoles with a positive vertical component and horizontal component along the positive *x*-axis, located at -2 *m* and not less than -4 *m* of depth, respectively, lead to admit the existence of a vertical system of tombs. The topmost tomb is very likely limited in the depth range from -1 *m* to -2.5 *m*, where the second feature is fully included.

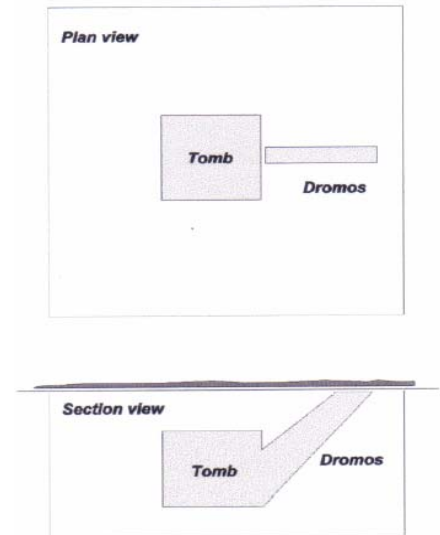

**Figure 9** Plan view and cross-section of the standard model of *dromos-chamber* tomb in the Sabine Necropolis at Colle del Forno, Rome, Italy.

The dipole is in this case an equivalent source, very useful to collocate the centre of the lacking magnetized masses. However, from the physical point of view, the hypothesized tombs can be better described by the JOP tomography of fig.11. In fact, continuous sequences of non-dissipative microscopic Ampère current elements must appear around the walls of the tombs, describing macroscopic concentric current loops lying in planes perpendicular to the axis of the equivalent dipoles. The tomography in figs.11a and 11b clearly show the traces of the dominant central loops, which can ideally be drawn with a line through the highest absolute values of the $\eta_{zx}^{(J)}$ and $\eta_{zy}^{(J)}$ functions.

### *Application to volcanology*

An aeromagnetic total field survey was done in 1978 by the Italian Oil Company AGIP to study the volcano-geothermal structure of Mount Vesuvius (see fig.12 for site location). A cesium optical pumping magnetometer was used at a constant height of 1460 *m* asl. Vesuvius is considered as one of the most risky active volcanoes in the world because of its closeness to the city of Naples, Italy. Results achieved so far (Iuliano, Mauriello and Patella 2001, and references therein) indicate that the shallow part of the volcano is made of a unique central plumbing system, entirely filled of altered volcanics in the summit portion. This peculiarity makes the volcano extremely hazardous, since explosion becomes a highly probable event in case of renewal of extruding activity. A magnetic study can notably help to ascertain whether such condition really occurs. To this purpose, we show the results from the application of the 3D probability tomography to the aeromagnetic total field map above Vesuvius (after Cassano and La Torre, 1987), depicted in the top slice of fig.13a and fig.14a. Subtracting the constant regional field in the whole survey area, with intensity 30,100 $nT$, inclination 56.30° and declination -1°, the residual effect is a large and intense positive anomaly with a peak value of approximately 1800 $nT$, located in the central part of the volcanic system.

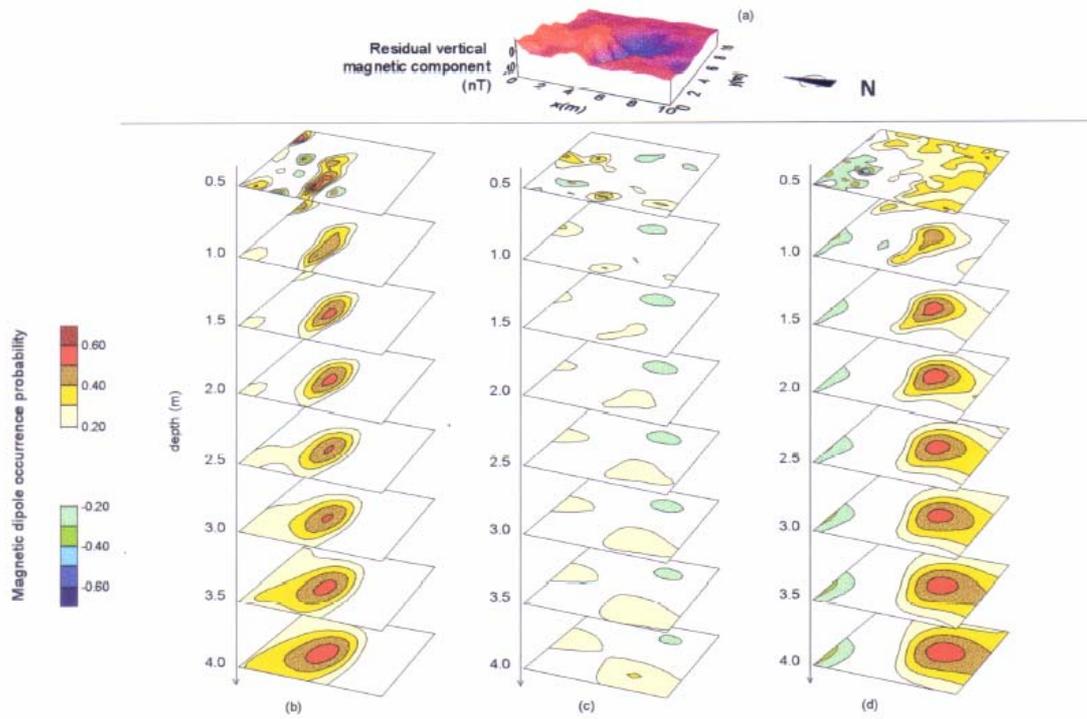

**Figure 10**   Magnetic probability tomography of a Sabine tomb at Colle del Forno, Rome, Italy. Experimental surface map of the *z*-component of a residual magnetic field (a) and 3D imaging of the magnetization occurrence probability function along the *x*-axis (b), *y*-axis (c) and *z*-axis (d).

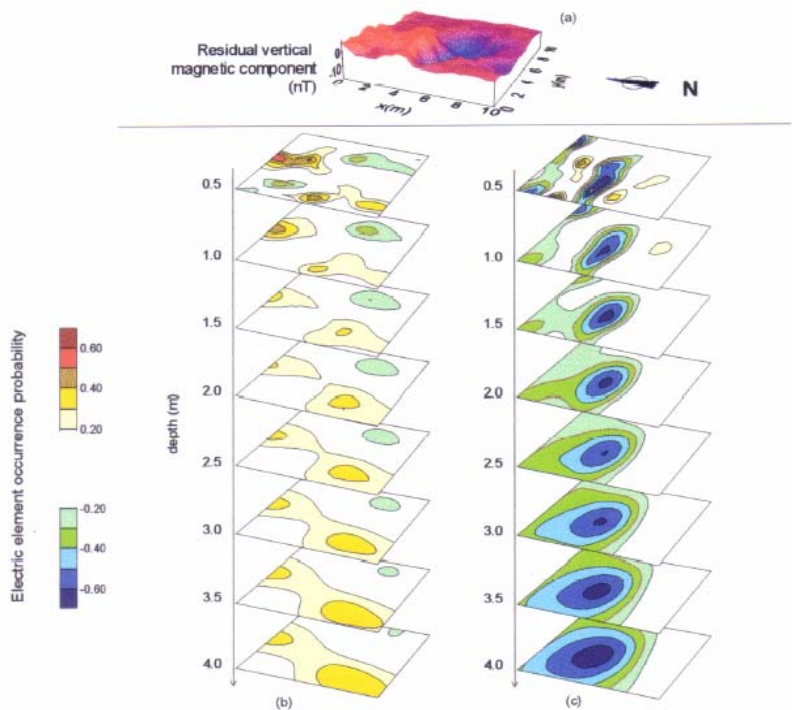

**Figure 11**   Magnetic probability tomography of a Sabine tomb at Colle del Forno, Rome, Italy. Experimental surface map of the *z*-component of the residual magnetic field (a) and. 3D imaging of the electric current occurrence probability function along the *x*-axis (b) and *y*-axis (c).

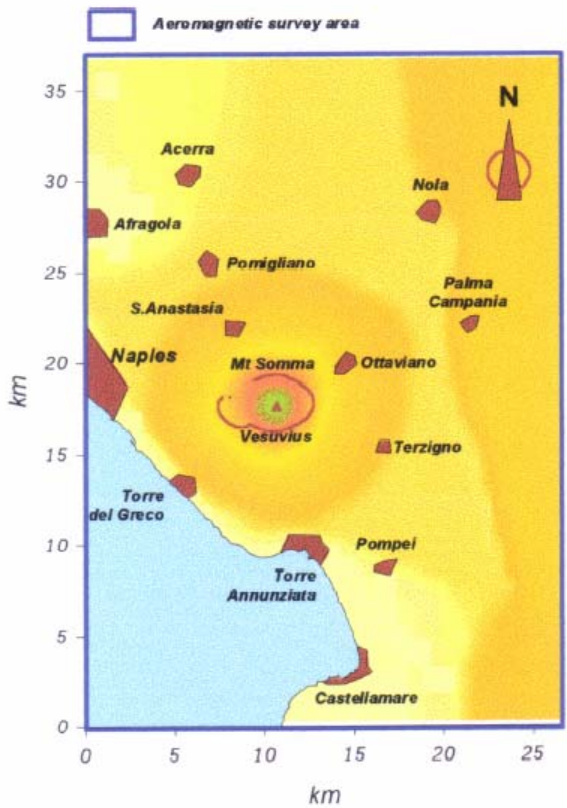

**Figure 12** The volcanic area of Vesuvius, Naples, Italy, with indication of the aeromagnetic survey area.

The slices in fig.13b,c,d show the behaviour of the MOP functions $\eta_{ux}^{(M)}$, $\eta_{uy}^{(M)}$ and $\eta_{uz}^{(M)}$, respectively. The slices in figs.14b,c,d show, instead, the behaviour of the JOP functions $\eta_{ux}^{(J)}$, $\eta_{uy}^{(J)}$ and $\eta_{uz}^{(J)}$, respectively. We observe again a close similarity with the tomographies relative to the inclined dipole, depicted in fig.7 and fig.8. Hence, a magnetic dipole placed at about 2 $km$ of depth bsl, having a negative vertical component and a horizontal component placed along the positive $y$-axis, must be invoked to interpret the magnetic map above

Vesuvius. This leads to hypothesize the existence of a magnetised material completely filling the bowl-shaped top terminal part of the volcanic central conduit, thus confirming the model deducted from the interpretation of other geophysical datasets.

## 2D SOURCE ANALYSIS

We now develop the probability tomography theory in the case of a 2D structure. We assume that the strike direction coincides with the $y$-axis and the magnetic field has been measured on a generally irregular profile $\ell$, entirely lying in a vertical plane perpendicular to strike, *i.e.* parallel to the $x$-axis.

### 2D J-occurrence probability

Referring to eq.9 and considering a mesh with $Q$ nodes in the $(x,z)$-plane, the $\mathbf{B}(\mathbf{r})$ field due to $Q$ elementary infinite cylinders placed parallel to strike and with their axes through the nodes is written as

$$\mathbf{B}(\mathbf{r}) = \sum_{q=1}^{Q} \mathbf{\Pi}_q \times \int_{-\infty}^{+z} \frac{(\mathbf{r} - \mathbf{r}_q)}{\left|\mathbf{r} - \mathbf{r}_q\right|^3} dy_q , \qquad (32)$$

where $\mathbf{\Pi}_q$ is the source strength $\mathbf{P}_q$, given by eq.9, per unit of length of the $q$-th cylinder.

As in the 3D case, we consider the case in which a component of $\mathbf{B}(\mathbf{r})$ along a generic direction, identified by the unit vector $\mathbf{u}$, has been measured at different stations along $\ell$. We start again from the definition of signal power associated with $B_u(\mathbf{r})$ along $\ell$, given as

$$\lambda_u = \int_{(\ell)} B_u^2(\mathbf{r}) d\ell , \qquad (33)$$

which, using eq.32, can be explicitly written as

$$\lambda_u = \sum_{q=1}^{Q} \left[ \sum_{v=x,y,z} \Pi_{qv} \int_{(\ell)} B_{uv}(\mathbf{r}) \mathfrak{I}_{uv}(x - x_q, z - z_q) d\ell \right]. \qquad (34)$$

The scanner $\mathfrak{I}_{uv}(x-x_q,z-z_q)$ has components given by

$$\mathfrak{I}_{ux}(x - x_q, z - z_q) = \frac{2(z_q - z) \mathbf{j} \cdot \mathbf{u}}{(x - x_q)^2 + (z - z_q)^2} , \qquad (35a)$$

$$\mathfrak{I}_{uy}(x - x_q, z - z_q) = \frac{2[(z - z_q)\mathbf{i} \cdot \mathbf{u} - (x - x_q)\mathbf{k} \cdot \mathbf{u}]}{(x - x_q)^2 + (z - z_q)^2} , \qquad (35b)$$

$$\mathfrak{I}_{uz}(x - x_q, z - z_q) = \frac{2(x - x_q) \mathbf{j} \cdot \mathbf{u}}{(x - x_q)^2 + (z - z_q)^2} . \qquad (35c)$$

Assuming that the projection of $\ell$ onto the $x$-axis is a line segment of length $2X$, and applying Schwarz's

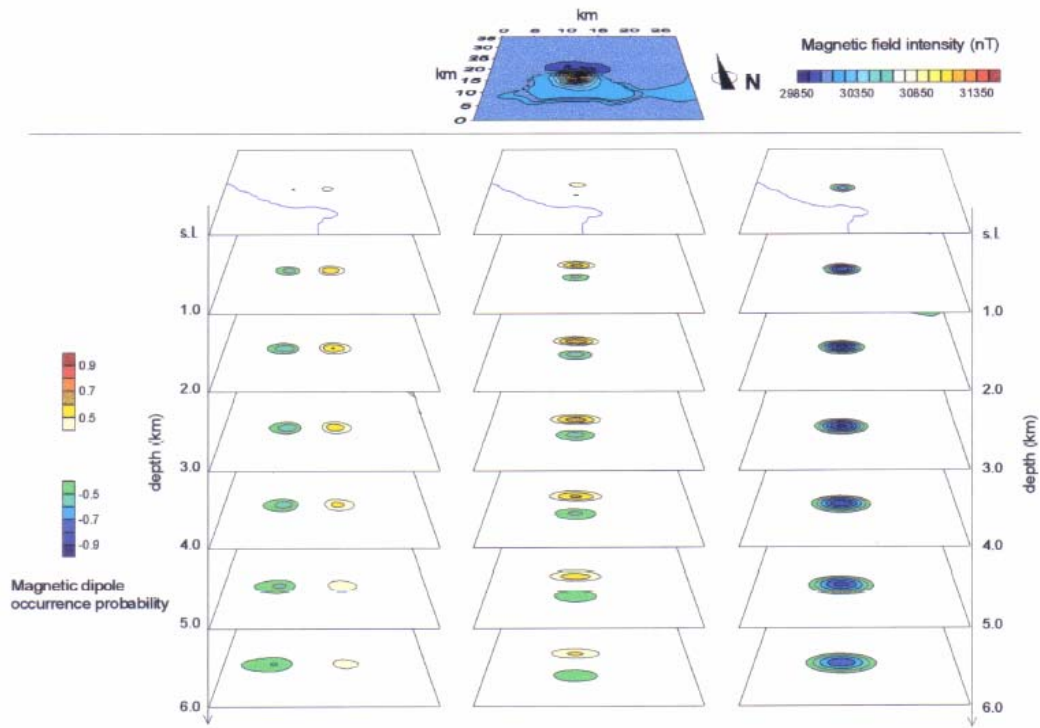

**Figure 13**    Magnetic probability tomography of Vesuvius volcano, Naples, Italy. Experimental surface map of the residual total magnetic field (a), and 3D imaging of the magnetization occurrence probability function along the *x*-axis (b), *y*-axis (c) and *z*-axis (d).

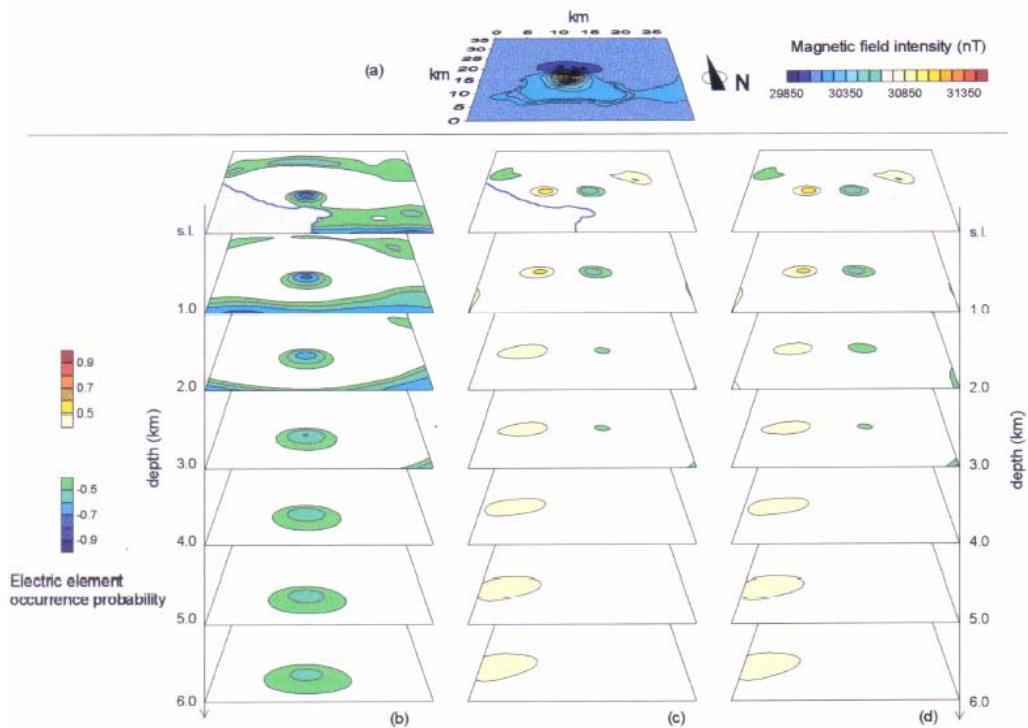

**Figure 14**    Magnetic probability tomography of Vesuvius volcano, Naples, Italy. Experimental surface map of the residual total magnetic field (a), and 3D imaging of the electric current occurrence probability function along the *x*-axis (b), *y*-axis (c) and *z*-axis (d).

 

bounding inequality to a generic integral in eq.34, we can at last define the 2D JOP function as

$$\eta_{uv}^{(J)}(x_q, z_q) = C_{uv}^{(J)} \int_{-X}^{X} B_u(x,z) \mathfrak{I}_{uv}(x-x_q, z-z_q) g_x(z) dx ,$$

$(v=x,y,z),$  (36)

where $g_x(z)$ is defined as the *x-ward topographic profile regularization* factor, given by

$$g_x(z) = \sqrt{1 + (\partial z/\partial x)^2} ,$$  (37)

and

$$C_{uv}^{(J)} = \left[ \int_{-X}^{X} B_u^2(x,z) g(z) dx \int_{-X}^{X} \mathfrak{I}_{uv}^2(x-x_q, z-z_q) g_x(z) dx \right]^{-1/2} ,$$

$(v=x,y,z).$  (38)

### 2D M-occurrence probability

Referring to eq.24, $\mathbf{B}(\mathbf{r})$ can be discretized as follows

$$\mathbf{B}(\mathbf{r}) = \sum_{q=1}^{Q} \int_{-\infty}^{\infty} \frac{3\mathbf{n}_q (\mathbf{n}_q \cdot \boldsymbol{\delta}_q) - \boldsymbol{\delta}_q}{|\mathbf{r} - \mathbf{r}_q|^3} dy_q ,$$  (39)

where $\boldsymbol{\delta}_q$ is the magnetic moment $\mathbf{d}_q$ per unit of length of the $q$-th cylinder.

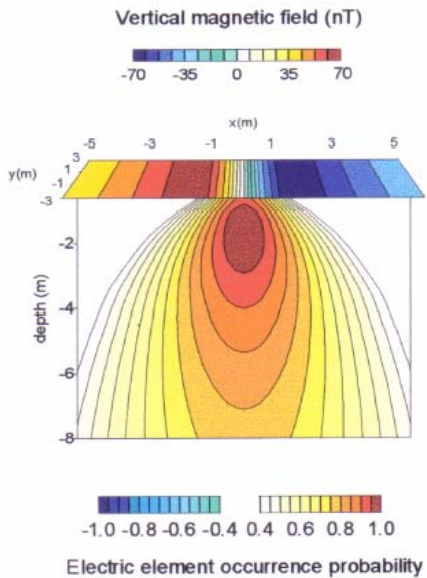

Vertical magnetic field (nT)

Electric element occurrence probability

**Figure 15** Probability tomography for a horizontal current wire in the case of measurement of the *z*-component of the magnetic field. Synthetic surface map of $B_z$ (a), and 3D imaging of the electric current occurrence probability function along the *y*-axis (b).

Following the same procedure as before, we define the 2D MOP function as

$$\eta_{uv}^{(M)}(x_q, z_q) = C_{uv}^{(M)} \int_{-X}^{X} B_u(x,z) \mathfrak{R}_{uv}(x-x_q, z-z_q) g_x(z) dx ,$$

$(v=x,y,z),$  (40)

where the explicit expressions of the three functions $\mathfrak{R}_{uv}(x-x_q, z-z_q)$, $(v=x,y,z)$, are given by

$$\mathfrak{R}_{ux}(x-x_q, z-z_q) =$$  (41a)

$$= \frac{2\{[(x-x_q)^2 + (z-z_q)^2] \mathbf{i} \cdot \mathbf{u} + 2(x-x_q)(z-z_q) \mathbf{k} \cdot \mathbf{u}\}}{[(x-x_q)^2 + (z-z_q)^2]^2} ,$$

$$\mathfrak{R}_{uy}(x-x_q, z-z_q) = 0 ,$$  (41b)

$$\mathfrak{R}_{uz}(x-x_q, z-z_q) =$$  (41c)

$$= \frac{2\{[(z-z_q)^2 - (x-x_q)^2] \mathbf{k} \cdot \mathbf{u} + 2(x-x_q)(z-z_q) \mathbf{i} \cdot \mathbf{u}\}}{[(x-x_q)^2 + (z-z_q)^2]^2} ,$$

and

$$C_{uv}^{(M)} = \left[ \int_{-X}^{X} B_u^2(x,z) g_x(z) dx \int_{-X}^{X} \mathfrak{R}_{uv}^2(x-x_q, z-z_q) g_x(z) dx \right]^{-1/2} ,$$

$(v=x,y,z).$  (42)

### A 2D SYNTHETIC EXAMPLE

In order to test the resolution power of the probability tomography also in the 2D case, we show the results of a synthetic case. Consider an infinitely long horizontal wire, parallel to the *y*-axis of a given reference system, at -1.5 *m* of depth below a flat ground surface, carrying a steady current of 1 *A* in the positive direction of the *y*-axis. We analyze the case in which the *z*-component of $\mathbf{B}(\mathbf{r})$, say $B_z(\mathbf{r})$, is supposed to have been measured.

By eq.35a,b,c and eq.41a,b,c, the three components of the $\mathfrak{I}_{zv}$ and $\mathfrak{R}_{zv}$ functions are given respectively as

$$\mathfrak{I}_{zx}(x-x_q, z-z_q) = 0 ,$$  (43a)

$$\mathfrak{I}_{zy}(x-x_q, z-z_q) = \frac{2(x_q - x)}{(x-x_q)^2 + (z-z_q)^2} ,$$  (43b)

$$\mathfrak{I}_{zz}(x-x_q, z-z_q) = 0 ,$$  (43c)

and

$$\mathfrak{R}_{zx}(x-x_q, z-z_q) = \frac{4(x-x_q)(z-z_q)}{[(x-x_q)^2 + (z-z_q)^2]^2} ,$$  (44a)

$$\mathfrak{R}_{zy}(x-x_q, z-z_q) = 0 ,$$  (44b)

$$\mathfrak{R}_{zz}(x-x_q, z-z_q) = \frac{2[(z-z_q)^2 - (x-x_q)^2]}{[(x-x_q)^2 + (z-z_q)^2]^2} .$$  (44c)

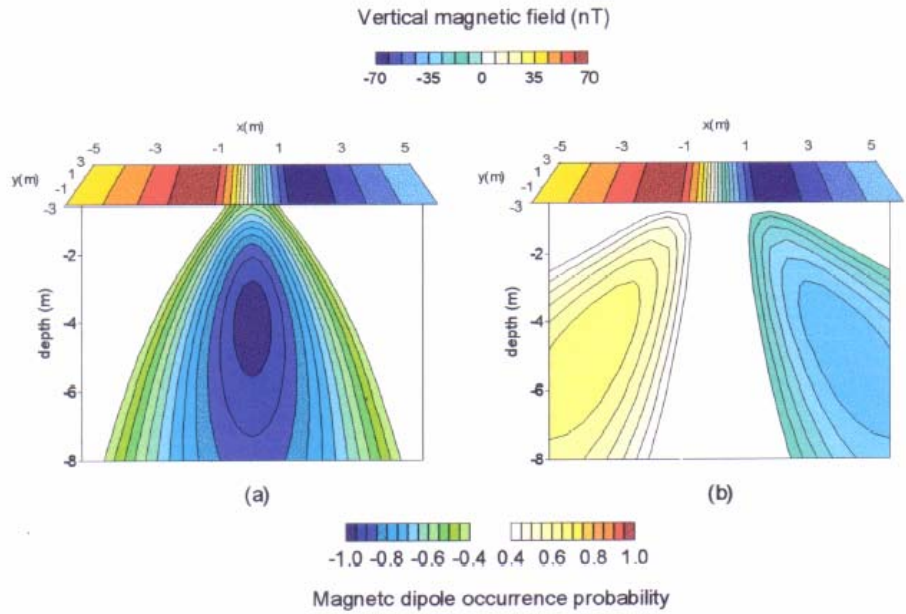

**Figure 16**  Probability tomography for a horizontal current wire in the case of measurement of the *z*-component of the magnetic field. Synthetic surface map of $B_z$ (a), and 3D imaging of the magnetization occurrence probability function along the *x*-axis (b) and *z*-axis (c).

The top horizontal slice in fig.15 and figs.16a,b shows the simulated $B_z$ field map. The vertical sections in fig.15 depicts the behaviour of the only computable

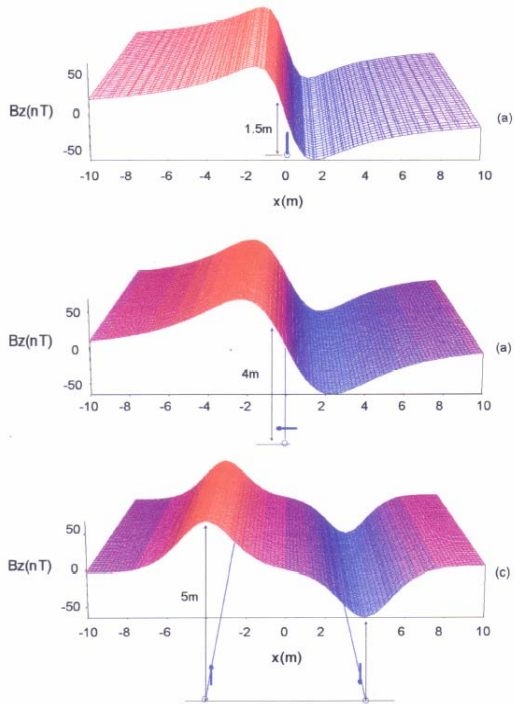

**Figure 17**  Response of dipole distributions equivalent to the horizontal current wire model. The $B_z$ map due to the original horizontal wire (a), to a set of parallel horizontal magnetization vectors (b) and to two sets of parallel vertical magnetization vectors with opposite orientation (c).

JOP function $\eta_{zy}^{(J)}$ (see eq.43a,b,c). Fig.16a and fig.16b show, instead, the 2D behaviour of the MOP functions $\eta_{zx}^{(M)}$ and $\eta_{zz}^{(M)}$, respectively (see eq.44a,b,c).

The JOP tomographic vertical section of fig.15 is very neat: it shows a positive nucleus, whose maximum value is located at -1.5 *m* of depth, in correspondence of the original current line source.

The MOP tomography of fig.16 requires, instead, a deeper analysis due to its greater complexity. At first, a negative nucleus appears in fig.16a at -4 *m* of depth, where the lowest $\eta_{zx}^{(M)}$ value occurs. It represents the trace of a strip of magnetization vectors oriented along the negative *x*-axis, providing an equivalent model with equally high occurrence probability. The $B_z$ map, due to the original current wire, redrawn in fig.17a, can in fact be accurately reproduced using a transverse magnetized strip, as shown in fig.17b. Moreover, two nuclei with opposite sign appear in fig.16b with maximum absolute values of $\eta_{zz}^{(M)}$ around -5 *m* of depth. This new pair of nuclei, though showing a notably lower occurrence probability, represent the trace of two strips of vertical magnetization vectors with opposite orientation, which can still explain the $B_z$ map of fig.17a, as illustrated in fig.17c.

## CONCLUSION

Currently used magnetic source localization procedures can be grouped wihin a rigid deterministic framework, despite the incomplete nature of the whole observation process. In order to avoid this current attitude, we have presented in this paper the principles of a new source localization method, based on a probabilistic approach. The new method, which of course does not require any change in data acquisition strategy, is quite different from all the previous source localization methods also because it does not require any *a priori* information to prime the inversion process. The new method strictly deals with the pure physical aspects of the secondary magnetic field created within earth's structures, without imposing any geological or other exotic constraints. In this frame, occurrence probability functions of electric currents and magnetic dipoles have been derived as the propermost entities to image a localization pattern of the sources of the magnetic anomalies observed above the ground, directly related to the intrinsic resolution power of the magnetic geophysical exploration method.